\documentclass[12pt]{article}
\usepackage{latexsym}
\newcommand{\xxx}[1]{}
\newcommand{\mysection}[1]{\section{#1}
   \hspace{0.8cm}\setcounter{equation}{0}}
\renewcommand{\theequation}{\arabic{section}.\arabic{equation}}
\newcommand{\myappendix}{\appendix
   \renewcommand{\theequation}{\Alph{section}.\arabic{equation}}
   \vspace{30pt} \noindent {\Large \bf Appendices}}
\def\thefootnote{\fnsymbol{footnote}}
\newcommand{\vev}{{\it vev}}
\def\tG{{\tilde G}}
\def\tc{{\tilde c}}

\def\tk{{\tilde k}}

\def\ts{{\tilde s}}
\def\tL{{\tilde L}}

\def\hE{{\hat E}}

\def\hO{{\hat\Omega}}
\def\tK{{\tilde K}}

\def\hR{{\hat R}}

\def\hL{\hat{L}}

\def\hS{\hat{S}}

\def\hK{\hat{K}}

\def\hb{\hat{b}}

\def\dx{\delta_X}
\def\Gx{G_X}
\def\qa{{q_A/q}}

\def\[{\left [}
\def\]{\right ]}
\def\({\left (}
\def\){\right )}

\def\pp{\partial}

\def\bTh{\bar{\Theta}}
\def\ux{$U(1)_X$}
\def\ua{$U(1)_a$}

\def\T{\bar{T}}
\def\S{\bar{S}}

\def\re{{\rm Re}}
\newcommand{\beq}{\begin{equation}}
\newcommand{\eeq}{\end{equation}}
\newcommand{\bea}{\begin{eqnarray}}
\newcommand{\eea}{\end{eqnarray}}

\def\G{{\cal G}}

\def\L{{\cal L}}
\newcommand{\superint}{\int \diff^{4}\theta \, }

\newcommand{\diff}{\mbox{d}}

\def\D{{\cal D}}
\def\bD{{\bar{\cal D}}}

\newcommand{\DbDb}{{\bar{\cal D}}^2}

\newcommand{\Dbadd}{\bar {\cal D}_{{\dot \alpha}}}

\def\CA{|C_A|^2}

\def\vx{V_X}

\def\pp{\partial}
\def\rvev{\right\rangle}
\def\lvev{\left\langle}

\def\hel{\hat\ell}

\def\Del{\Delta}

\def\del{\delta}

\def\e{\epsilon}

\newcommand{\bbar}[1]{{\overline{#1}}}

\newcommand{\ord}[1]{{O(#1)}}
\newcommand{\myref}[1]{(\ref{#1})}
\newcommand{\beqa}{\begin{eqnarray}}
\newcommand{\eeqa}{\end{eqnarray}}
\newcommand{\nnn}{ \nonumber \\ }
\newcommand{\p}{\partial}
\newcommand{\hc}{{\rm h.c.}}
\newcommand{\W}{{\cal W}}
\newcommand{\Wb}{{\bbar{{\cal W}}}}
\newcommand{\chiproj}{(\bD^2 - 8R)}
\newcommand{\hchiproj}{(\hat \bD^2 - 8 \hat R)}
\newcommand{\bchiproj}{(\D^2 - 8 \bar R)}
\newcommand{\chiprojb}{(\D^2 - 8 \bar R)}

\newcommand{\fourth}{{1 \over 4}}
\newcommand{\uone}{$U(1)$}
\newcommand{\uones}{$U(1)$'s}

\newcommand{\myvev}[1]{{\langle #1 \rangle}}

\newcommand{\Dk}{\Delta k}
\newcommand{\third}{{1 \over 3}}
\newcommand{\tr}{\mathop{{\hbox{Tr} \, }}\nolimits}

\newcommand{\half}{{1 \over 2}}

\newcommand{\Az}{{A_0}}
\newcommand{\DA}{\Del_A}
\newcommand{\DB}{\Del_B}

\newcommand{\ddd}{\nnn && \quad}

\def\hSig{\hat\Sigma}
\newcommand{\adot}{{\dot \alpha}}

\begin{document}

\begin{titlepage} 
            
\hfill   LBNL-50951
            
\hfill   UCB-PTH-02/28
            
\hfill   hep-th/0206249

\hfill   June 2002

\begin{center}

\vspace{18pt}
{\bf \Large More Modular Invariant Anomalous ${\bf U(1)}$ 
Breaking}\footnote{This work was supported in part by the
Director, Office of Science, Office of High Energy and Nuclear
Physics, Division of High Energy Physics of the U.S. Department of
Energy under Contract DE-AC03-76SF00098 and in part by the National
Science Foundation under grant PHY-0098840.}

\vspace{18pt}
Mary K. Gaillard\footnote{E-Mail: {\tt MKGaillard@lbl.gov}}
{\em and} Joel Giedt\footnote{E-Mail: {\tt JTGiedt@lbl.gov}}

\vspace{18pt}

{\em Department of Physics, University of California \\
and \\ Theoretical Physics Group, Bldg. 50A5104,
Lawrence Berkeley National Laboratory \\ Berkeley, 
CA 94720 USA.}\\[.1in]

\vspace{18pt}

\end{center}

\begin{abstract}
We consider the case of several scalar fields, charged under a number
of $U(1)$ factors, acquiring vacuum expectation values due to an
anomalous $U(1)$.  We demonstrate how to make redefinitions at the
superfield level in order to account for tree-level exchange of vector
supermultiplets in the effective supergravity theory of the light
fields in the supersymmetric vacuum phase.  Our approach builds upon
previous results that we obtained in a more elementary case.  We find
that the modular weights of light fields are typically shifted from
their original values, allowing an interpretation in terms of the
preservation of modular invariance in the effective theory.  We
address various subtleties in defining unitary gauge that are
associated with the noncanonical K\"ahler potential of modular
invariant supergravity, the vacuum degeneracy, and the role of the
dilaton field.  We discuss the effective superpotential for the light
fields and note how proton decay operators may be obtained when the
heavy fields are integrated out of the theory at the tree-level.  We
also address how our formalism may be extended to describe the
generalized Green-Schwarz mechanism for multiple anomalous $U(1)$'s
that occur in four-dimensional Type I and Type IIB string
constructions.
\end{abstract}

\end{titlepage}

\newpage
\renewcommand{\thepage}{\roman{page}}
\setcounter{page}{2}
\mbox{ }

\vskip 1in

\begin{center}
{\bf Disclaimer}
\end{center}

\vskip .2in

\begin{scriptsize}
\begin{quotation}
This document was prepared as an account of work sponsored by the United
States Government. Neither the United States Government nor any agency
thereof, nor The Regents of the University of California, nor any of their
employees, makes any warranty, express or implied, or assumes any legal
liability or responsibility for the accuracy, completeness, or usefulness
of any information, apparatus, product, or process disclosed, or represents
that its use would not infringe privately owned rights. Reference herein
to any specific commercial products process, or service by its trade name,
trademark, manufacturer, or otherwise, does not necessarily constitute or
imply its endorsement, recommendation, or favoring by the United States
Government or any agency thereof, or The Regents of the University of
California. The views and opinions of authors expressed herein do not
necessarily state or reflect those of the United States Government or any
agency thereof of The Regents of the University of California and shall
not be used for advertising or product endorsement purposes.
\end{quotation}
\end{scriptsize}

\vskip 2in

\begin{center}
\begin{small}
{\it Lawrence Berkeley Laboratory is an equal opportunity employer.}
\end{small}
\end{center}

\newpage
\renewcommand{\thepage}{\arabic{page}}
\setcounter{page}{1}
\def\thefootnote{\arabic{footnote}}
\setcounter{footnote}{0}

\mysection{Introduction}
In our previous article \cite{GG02a} we studied the effective
supergravity theory obtained in the presence of
an anomalous \uone, that for the remainder of
this article we will denote \ux.  In the simple
case that we investigated, a single scalar field
charged under \ux\ acquired a nonvanishing
vacuum expectation value ({\it vev}).  The associated
chiral multiplet was ``eaten'' by the \ux\ vector
multiplet to form a massive vector multiplet.
We eliminated tree-level exchange of this massive
vector multiplet by redefinitions that eliminated
linear couplings between the heavy and light fields.
We demonstrated that this redefinition can be
made at the superfield level, while maintaining
manifest modular invariance and the (modified) linearity
conditions,
\beq
\chiproj L = - \sum_a (\W \W)_a, \qquad
\bchiproj L = - \sum_a (\Wb \Wb)_a,
\label{abc}
\eeq
for the linear superfield $L$, whose
lowest component is the real scalar associated
with the dilaton.  A comparison with redefinitions
at the component field level provided assurances
that the superfield approach was reliable.

Our motivations stemmed from the prevalence of
a \ux\ factor in the string scale gauge group of
semi-realistic string compactifications; for example,
in a recent study \cite{Gie02a} of a certain
class of standard-like heterotic $Z_3$ orbifold
models, it was found that 168 of 175 models
had an anomalous \ux.  Clearly the simple case
considered in our previous article does not
address the complications that arise in the
semi-realistic models that we seek to understand,
since the scalars that get \vev's due to the
\ux\ are typically charged under several \uone\
factors and multiple scalars must generally get
\vev's in order for the {\it D-terms} of the several
$U(1)$'s to (approximately) vanish.  The purpose
of this paper is to examine the effective
supergravity theory when these two generalizations
are made, for the case of the supersymmetric
vacuum phase.  As has been noted previously,
the supersymmetric vacuum is approximately the
stable vacuum in the case where dynamical supersymmetry
breaking {\it via} gaugino condensation occurs 
in an effective supergravity context \cite{Bar98}.
Thus the scenario studied here represents a
{\it bona fide} starting point for the
effective supergravity theory of semi-realistic
models with a \ux.

We start with the effective theory at the string scale defined
as in \cite{GG02a}:
\beq
\L = \superint\tL + \L_Q + \L_{th},
\label{tlg}
\eeq
where $\tL$ is the real superfield functional
\beq
\tL = E\[- 3 + 2Ls(L) + L\(bG - \dx\vx\)\]
= E\[- 3 + 2LS\],
\label{oit}
\eeq
and where the K\"ahler potential given by
\beqa
K &=& k(L) + G + \sum_Ae^{G^A + 2 \sum_a q_A^a V^a}|\Phi^A|^2, \quad 
G = \sum_Ig^I, \quad G^A = \sum_Iq^A_Ig^I, 
\nnn
g^I &=& -\ln(T^I + \T^I), \quad k(L) = \ln L + g(L).
\label{kpt}
\eeqa
In the dual chiral formulation $s(L) \to {\rm Re}s;\;\lvev s(L)\rvev = g^{-2}$ at the 
string scale.  Canonical normalization of the Einstein term requires:
\beq 
k'(L) = - 2Ls'(L).
\label{canon}
\eeq
For reasons explained in \cite{GG02a}, 
we are not using $U(1)_K$ superspace for 
the Abelian gauge groups that are broken at
the string scale by the anomalous \ux.  

Since the underlying
theory is anomaly free, it is known \cite{UXR}
that the apparent anomaly is canceled by 
a four-dimensional version of the
Green-Schwarz (GS) mechanism \cite{GS84}.  This
leads to a Fayet-Illiopoulos (FI) term in
the effective supergravity Lagrangian.  Ignoring
nonperturbative corrections to the
dilaton K\"ahler potential,
\beq
D_X = \sum_A {\p K \over \p \phi^A} q^X_A \phi^A + \xi,
\qquad \xi = {g_s^2 \tr T_X \over 192 \pi^2} m_P^2,
\label{eq1}
\eeq
where $K$ is the K\"ahler potential, $q_A^X$ is the
\ux\ charge of the scalar matter field $\phi^A$,
$\xi$ is the FI term, $T_X$ is the charge generator
of \ux, $g_s$ is the unified (string scale) gauge coupling,
and $m_P=1/\sqrt{8\pi G} = 2.44 \times 10^{18}$ GeV is
the reduced Planck mass.  In the remainder we
work in units where $m_P=1$.

Up to perturbative loop effects,
the chiral dilaton formulation has $g_s^2 = 1/\re \myvev{s}$,
where $s= S|$ is the lowest component of the
chiral dilaton superfield $S$.  However, once
higher order and nonperturbative corrections
are taken into account the chiral dilaton
formulation becomes inconvenient.  The dual
linear multiplet formulation---that relates
a (modified) linear superfield $L$ to $\{ S,\bar S\}$
through a duality transformation---provides
a more convenient arrangement of superfield degrees
of freedom due to the neutrality of $L$
with respect to target-space duality transformations
(hereafter called {\it modular transformations}).
In the limit of vanishing nonperturbative
corrections to the dilaton K\"ahler potential,
$g_s^2 = 2 \myvev{\ell}$, where $\ell = L|$.
Throughout this article we use the linear multiplet
formulation \cite{LMULT,BGG89}.  Except where noted above, we
use the $U(1)_K$ superspace formalism \cite{Mul86,BGG89,BGG01}.
(For a review of the $U(1)_K$ superspace
formalism see \cite{BGG01}; for a review of
the linear multiplet formulation see \cite{GG99}.)

In the linear multiplet formulation,
including nonperturbative corrections to the dilaton
K\"ahler potential, the FI term becomes
\beq
\xi(\ell) = {2 \ell \tr T_X \over 192 \pi^2}.
\label{eq2}
\eeq
Consequently, the background dependence of the FI
term in \myref{eq2} arises from $\myvev{\ell}=\myvev{L|}$.
The FI term induces nonvanishing
{\it vev}'s for
some scalars $\phi^A$ as the scalar potential
drives $\myvev{D_X} \to 0$, if supersymmetry is unbroken.
The nonvanishing \vev's in the
supersymmetric vacuum phase can be related to the
FI term.  Then
$\myvev{L|}$ serves as an order parameter for the vacuum
and all nontrivial \vev's can be written as some
fraction of $\myvev{L|}$.  {\it Our approach in
what follows will be to promote this to a
superfield redefinition.}
Thus we impose the superfield identity 
\beq 
\( {\pp K\over\pp V_a} + 2L{\pp S\over\pp V_a} \)_{\DA=0} = 
\( {\pp K\over\pp V_a} \)_{\DA=0} - L\dx\del_{Xa}  = 0,\label{sfid}
\eeq
where $\DA$ are superfields, to be defined below,
that vanish in the supersymmetric vacuum.
This assures vanishing of 
the D-terms at the \ua\, symmetry breaking
scale while maintaining manifest supersymmetry below that scale.  The
latter point was demonstrated in detail, at both the superfield and the
component field levels, for the toy model studied in~\cite{GG02a}.

$\L_Q$ is the quantum correction~\cite{gt,kl,gnw} 
that contains the field theory anomalies canceled by the GS terms:  
\bea
\L_Q &=& - \int d^4\theta
{E\over 8R}\sum_a \W_a^\alpha P_\chi B_a
\W_\alpha^a + {\rm h.c.},
\label{lqa} \\
B_a(L,V_X,g^I) &=& \sum_I (b-b^I_a) g^I - \dx\vx + f_a(L),
\label{lqb}
\eea
where $P_\chi$ is the chiral projection operator~\cite{1001}: 
$P_\chi \W^\alpha = \W^\alpha$, that reduces
in the flat space limit to $(16\Box)^{-1}\bD^2\D^2$, and
the $L$-dependent piece $f_a(L)$ is the ``2-loop'' contribution~\cite{gt}.
The string-loop contribution is~\cite{thresh}
\beq
\L_{th} = -\superint\,{E\over 8R}
\sum_{a,I} b_a^I (\W \W)_a
\ln\eta^2(T^I) + {\rm h.c.} 
\label{lth}
\eeq
For each $\Phi^A$, the \ux\ charge is denoted 
$q_A^X$ while $q^A_I$ are the modular weights.  
The conventions chosen here imply \ux\ gauge 
invariance under the transformation
\beq 
\vx \to  \vx' = \vx + {1\over2}\(\Theta + \bTh\), 
\quad \Phi^A \to \Phi'^A = e^{-q_A^X \Theta}\Phi^A.
\label{baa}
\eeq
The GS coefficients $b$ and $\dx$
must be chosen to cancel the quantum field
anomalies under modular and \ux\ transformations
that would be present in the absence of
the GS counterterms \cite{UXR,gsterm}.
It is not hard to check that the correct
choices are given by:
\beqa
\dx &=& -{1\over2\pi^2}\sum_AC^A_{a\ne X}q_A^X
= -{1\over48\pi^2} \tr T_X ,
\label{dxnorm} \\
8 \pi^2 b & = & 8 \pi^2 b_a^I+ C_a - \sum_A(1-2q^A_I)C_a^A.
\label{bdefs}
\eeqa

In Section \ref{gfc} we address complications
introduced by the occurrence of several chiral superfields
in the theory, some with scalar components getting
\vev's and others whose scalar components do not
get \vev's.  Unitary gauge and the decomposition of
chiral multiplets into light and heavy multiplets is
addressed.
In Section \ref{ggg} we discuss the case where
the fields getting large \vev's due to the \ux\ are
charged under several \uones.  It is shown how
the ``eating'' of chiral multiplets proceeds in this
situation.  Further, we look into the reinterpretation of
the modified linearity conditions when written
in terms of vector superfields with vanishing \vev's
after the necessary Weyl transformation is made.
In Section \ref{spo} we investigate the effective
superpotential that results when the field redefinitions
are made.  We demonstrate how new operators are
obtained when tree-level exchange is accounted for.
As an example of the relevance of such effective
operators, we note how proton decay 
operators may be obtained.
In Section \ref{con} we present our conclusions
and lay out items that remain to be investigated.
In Appendix \ref{ecd} we describe the necessary condition
to have a canonical Einstein term.
In Appendix \ref{wtr} we formulate the Weyl transformation
necessary to eliminate linear couplings to the heavy
fields, maintain the linearity of $L$ and preserve
the canonical normalization of the Einstein term.
In Appendix \ref{usi} we extend our formalism to
the generalized GS cancellation of several anomalous
\uones\ that occurs in Type IIB and Type I string
theories.  Here, several linear multiplets are involved
that do not correspond to the dilaton.


\mysection{Generalized Field Content}
\label{gfc}
In this section we assume a set of chiral superfields $\Phi^A$
that are charged only under \ux.  Then the first equality
in \myref{kpt} is simply
\beq 
K = k(L) + G + \sum_A K_{(A)}, \qquad
K_{(A)}=e^{G^A + 2 q_A V}|\Phi^A|^2.
\label{bab}
\eeq 
We will encounter subtleties in going to unitary gauge, in part
because the sum in \myref{bab} is weighted by $\exp G^A$, which may be
different for the various matter fields getting \vev's to cancel the
FI term.  Our emphasis, as in our previous article, will be on keeping
modular invariance manifest.  We parameterize the vacuum in terms of
modular invariant \vev's $\myvev{K_{(A)}}$, which by the
vanishing of \myref{eq1} in the supersymmetric vacuum we can rewrite
in terms of $\ell$.  We promote this to a superfield redefinition
throughout, in order to retain explicit supersymmetry, as well as
modular invariance, in the effective theory below the $V$ mass
scale.  When there is more than one scalar \vev\, contributing to the
\ux\, gauge symmetry breaking there are further difficulties in this
approach, as discussed in Section \ref{multvevs}, that are related to the
degeneracy of the vacuum, and the role of the linear multiplet
as an order parameter.

\subsection{One Vev}
Suppose first that only one field 
$\Phi^\Az \equiv e^{\Theta}$, with $q_\Az = q$,
$q^\Az_I = q^I$, gets a \vev.  In this simple case,
unitary gauge is obtained as in our
previous study; under \myref{baa} we have
\beq 
V \to V' = V + {1\over2q}\(\Theta + \bTh\),
\qquad \Phi^\Az \to \Phi'^\Az = e^{-\Theta}\Phi^\Az = 1.
\label{gtr}
\eeq
The field $V'$ describes a massive vector multiplet 
in the unitary gauge; it has ``eaten'' the chiral
superfield $\Theta$ and its conjugate.  The contribution
to the K\"ahler potential from this field then
simplifies to ($G_q \equiv G^\Az$):
\beq
K_{(\Az)} = e^{G_q + 2qV'} .
\eeq
For the other chiral superfields we have from \myref{baa}
\beq 
\Phi^A \to \Phi'^A = e^{-q_A\Theta/q}\Phi^A
\eeq
and the corresponding contributions to the K\"ahler
potential become
\beq
K_{(A)} = e^{G^A + 2q_AV}|\Phi^A|^2 
= e^{G^A + 2q_AV'}|\Phi'^A|^2.
\eeq

Because $\myvev{\Phi^\Az} \not= 0$ we see that if we
assume\footnote{We can always do this by going
to WZ gauge for $V$.  The same situation is
{\it not} true for $V'$.}
$\myvev{V}=0$ we have after the gauge
transformation \myref{gtr} that $\myvev{V'} \not= 0$. 
Moreover, while $V$ is modular invariant, $V'$ is not,
as is obvious from \myref{gtr} since $\Theta$ is not
modular invariant.
It is convenient to work instead with a modular
invariant vector superfield with vanishing \vev;
 we therefore follow our previous approach and define
\beq
V' \equiv U + {1 \over 2q} \( \ln {\dx L \over 2q} - G_q \) ,
\qquad \myvev{U} \equiv 0.
\label{aad}
\eeq
Going over to this basis, we have
\beqa 
K_{(A)} &=& e^{G'^A + 2q_AU}\({\dx L\over2q}\)^{q_A/q}|\Phi'^A|^2, 
\label{kpp} \\
G'^A &=& \sum_Iq'^A_Ig^I, \qquad q'^A_I = q^A_I - q_Iq_A/q.
\label{smw}
\eeqa
In particular we have $q'^\Az_I=0$, yielding
\beq
K_{(\Az)} = e^{2 q U} {\dx L\over2q},
\label{kol}
\eeq
in agreement with Eq.~(31) of \cite{GG02a} for the
field that gets a \vev.
In the remainder of this subsection, $A \not= \Az$
as $\Phi^\Az$ has been ``eaten'' by the \ux\ superfield.

From the contributions \myref{kpp} to the K\"ahler potential,  
the {\it Einstein condition} (defined
by Eq.~\myref{aab} and the associated discussion
in Appendix~\ref{ecd}) is not satisfied in the primed
basis.  For instance, \myref{kol} was already shown
to lead to an $\ord{U}$ violation of the Einstein
condition in Eq.~(44) of \cite{GG02a}.  It therefore becomes
necessary to make field redefinitions that involve $L$
in order to repair this situation.  To preserve the
linearity of $L$, we accomplish this by a Weyl
transformation.  The details of this method are given
in Appendix~\ref{wtr}.  There we also show that the
identity \myref{sfid} assures the simultaneous 
elimination of linear
couplings to the heavy vector multiplet.

To proceed, it is convenient to work temporarily with the 
{\it modular invariant} real superfields 
\beq
X_A = e^{G'^A}\({\dx\over2q}\)^\qa|\Phi'^A|^2.
\label{yut}
\eeq
Since $A \not= \Az$, these fields all have vanishing
\vev's:  $\myvev{X_A}=0$.  We will exploit this, together with 
$\myvev{U}=0$,
to make expansions in $U$ and $X_A$, referred to
collectively as $\DA$ in the notation of Appendix~\ref{wtr}.
The redefinition \myref{yut} is a matter
of ``tidy bookkeeping,''
since it allows for coefficient functionals in Appendix~\ref{wtr}
that are functionals only of $L$ and not functionals of $g^I$.
It also ensures that the expansions made there are
modular invariant.

The K\"ahler potential \myref{bab} then becomes
\beq 
K = \tk(L) + G +{\dx L\over2q}\(e^{2 q U}-1\)
+ \sum_{A \not= \Az} L^\qa X_A e^{2q_A U}.
\label{bad}
\eeq
Furthermore in the basis $(L,g^I,\Del)$ the functional
in Eq.~\myref{oit} becomes
\beq
\tL =  E \[ -3 + 2LS(L,U,g^I) \], \qquad
S(L,U,g^I) = \ts(L) + \half \tG - {\dx \over 2} U,
\label{aac}
\eeq
where $S$ is the functional described
in Appendix \ref{ecd} and for convenience we define
\beqa
\ts(L) &=& s(L) - {\dx \over 4 q} \ln {\dx L \over 2 q},
\label{qw8} \\
\tilde G &=&
bG + {\dx \over 2q} G_q
= \sum_I \( b + {\dx \over 2q} q^I \) g^I.
\label{mqw}
\eeqa
We note that $\tk$ and $\ts$ are identically related by
the constraint \myref{tci}.

We next apply the results of Appendix~\ref{wtr}.
Note that \myref{bad} contains terms up to only
linear order in $X_A$; thus
the expansion \myref{wke} simplifies to
a power series in $U$:
\beqa
K = \tk(L) + G + k^U(L) U + \half k^{UU}(L) U^2
+ \sum_{A \not= \Az} X_A \[k^A(L) + k^{UA}(L) U\]
+ \ord{\Del^3}
\eeqa
with coefficients
\beq 
k^U(L) = \dx L, \qquad 
k^A(L) = L^\qa, \qquad
k^{UU}(L) = 2 q \dx L, \qquad 
k^{UA}(L) = 2 q_A L^{q_A/q}.
\eeq
A nice simplification also occurs for $S$:  we have
no $X_A$ terms and only a linear term in $U$,
with $s^U(L)= - \dx/2$ in the notation of
\myref{wse}.

The parameters of the Weyl transformation to
linear order are given by
\beq
\alpha^U(\hL) = \alpha_0(\hL), \qquad
\alpha^A(\hL) = {q_A  \over \dx q} 
\alpha_0(\hL) \hL^{(q_A-q)/q}.
\label{awq}
\eeq
The leading coefficients of the shifted K\"ahler potential $\hK$
and the functional $\hS$ are given by
$$
\hK^U = \hS^U = 0, \qquad 
\hK^{UU}(\hL) + 2 \hL \hS^{UU}(\hL) 
= 2q\dx\hL 
\( 1 + {\alpha_0(\hL) \over 6q} \),
$$
$$ 
\hK^A(\hL) = \(1 - {q_A\over q}\)\hL^\qa, 
\qquad 2\hL \hS^A(\hL) = {q_A\over q} \hL^\qa,
$$
$$
\hK^{UA}(\hL)+ 2 \hL \hS^{UA}(\hL) 
= 2 q_A \hL^\qa \( 1 + {\alpha_0(\hL) \over 6q} \),
$$
\beq
\hK^{AB}(\hL)+ 2 \hL \hS^{AB}(\hL) =
{q_A q_B \alpha_0(\hL) \over 3 q^2 \dx \hL}
\hL^{(q_A + q_B)/q} .
\eeq
The vanishing of $\hK^U$ and $\hS^U$ is a
consequence of the Weyl transformation, and
gives the desired result that linear couplings 
to the heavy multiplet $U$ be eliminated.  

The (naively) worrisome $U$ to $X_A$
couplings that arise from $\hK^{UA}(\hL)+ 2 \hL \hS^{UA}(\hL)$
are an artifact of our ``bookkeeping'' \myref{yut}.
These give $\ord{U |\Phi'^A|^2}$ contributions
in terms of the elementary superfields.  The tree exchange
of the heavy multiplet $U$ will give $\ord{|\Phi'^A|^2 |\Phi'^B|^2}$
effective terms involving space-time derivatives or auxiliary
components.  Since $\myvev{\Phi'^A} \equiv 0$ and auxiliary
component \vev's are of order of the gravitino mass, 
these are highly suppressed interaction terms
that are negligible for the purposes of our considerations.

Now we examine the content of the \ux\ field strength
in the new superfield basis.  The situation is very
similar to that which appeared in our previous 
article \cite{GG02a}.  Once we make the Weyl transformation,
spinorial derivatives are covariant with respect to
the new K\"ahler potential $\hK$, and the chiral
auxiliary superfield  $\hR$ is also defined with respect
the new $\hK$.  After the Weyl transformation we
have $\W_{V'} \to \widehat \W_{V'}$ wherever the \ux\ chiral
field strength appeared before, such as in the modified
linearity conditions \myref{abc} or the one-loop
effective terms \myref{lqa} and \myref{lth}.  However,
the redefinition \myref{aad} now leads to a reinterpretation
of this quantity:
\beqa
\widehat \W_{V'}^\alpha &=& - \fourth \hchiproj \hat \D^\alpha V' \nnn
&=& \widehat \W_U^\alpha -{1 \over 8q} \hchiproj \hat \D^\alpha
\( \ln {\dx \hL \over 2q} - G_q \)
-{1 \over 24 q} \hchiproj \hat \D^\alpha \Dk(\hL,\Del) . \quad
\label{kja}
\eeqa
In particular, the $\Del_A \equiv U$ term in $\Dk(\hL,\Del)$
gives additional contributions to the chiral field strength
for the new vector multiplet $U$.  At leading order in the
small quantities $\Del_A$, the correction yields
\beq
\widehat \W_{V'}^\alpha = \(1 + {\alpha^U \over 6q} \)
\widehat \W_U^\alpha + \cdots .
\eeq
Taking into account \myref{awq}, this result is in agreement
with Eq.~(79) of our previous work \cite{GG02a}.  A great
many other terms arise from the remainder of \myref{kja}.
However, these yield higher order terms in the effective
Lagrangian and are negligible at the order of analysis taken
up here.

\subsection{Multiple Vevs}
\label{multvevs}
Here the situation is generically more complicated and
it is consequently more difficult to decompose the
superfields into heavy and light subsets.
We first go to a {\it quasi-unitary gauge,} where
the meaning of this term will be made clear in
what follows.  Define
\beq
\lvev e^{G^A}|\Phi^A|^2\rvev = \CA,
\label{defca}
\eeq
where $C_A$ is a complex constant.  Writing, for $C_A\ne0$,
\beq \Phi^A = C_Ae^{\Theta^A}, \quad \Sigma^A = \Theta^A + \bTh^A
+ G^A, \quad \lvev\Sigma^A\rvev = 0,\eeq
we go to quasi-unitary gauge by making a gauge transformation
that eliminates the eaten chiral multiplet:
\bea 
V' &=& V + \Theta + \bTh, \quad \Phi'^A = e^{-2q_A\Theta}\Phi^A,
 \nonumber \\ \Theta &=& {1\over2Q}
\sum_Aq_AB_A\Theta^A, \quad Q = \sum_Aq^2_AB_A.\label{ugauge}
\eea
The coefficients $B_A$ are real constants.
$\Theta$ drops out of the Lagrangian written in terms of $V',\Phi'$.
If $C_A\ne0$ for $A = 1,\ldots,n$ we have for $A\le n$
\beq \Phi'^A = C_Ae^{\Theta'^A}, \quad
\Theta'^A = \Theta^A - 2q_A\Theta.\eeq
Only $n-1$ of the chiral fields $\Theta'^A$ are linearly independent:
\beq \sum_Aq_AB_A\Theta'^A = 0.\label{lindep}\eeq  
$\Theta$ is chiral but not modular invariant; $\Phi'^A$ is chiral with
modular weight
\beq q'^A_I = q^A_I - 2q_Aq_I, \quad q_I = {1\over2Q}\sum_Aq_AB_A
q_I^A.\label{modwts}\eeq
To obtain a modular invariant vector field we set
\bea U' &=& V' + \Gx = V + \Sigma, \quad \Sigma = \Theta + \bTh
+ \Gx, \nonumber \\ \Gx &=& {1\over2Q}\sum_Aq_AB_A G^A.\label{uprime}\eea
$U',\Sigma$ and the matter field contributions to the
K\"ahler potential
\beq K(\Phi) = \sum_AK_{(A)}, \quad K_{(A\le n)} = 
\CA e^{\Sigma'^A + 2q_AU'},\eeq
are modular invariant. The real fields
\beq \Sigma'^A = \Theta'^A +  \bTh'^A + G'^A, \quad G'^A = G^A
- 2q_AG_X,\eeq 
have vanishing $vev$'s and satisfy
\beq \sum_Aq_AB_A\Sigma'^A = 0. \label{gb}\eeq  
Finally we make a modular invariant shift in the vector field:
\beq 
U' = U + h(L) + \sum_Ab_A(L)\Sigma'^A, \quad\lvev U\rvev = 0,
\eeq
where $h,b_A$ 
are functionals of $L$ to be determined.  We require the D-term $vev$ to 
vanish:
\beq \sum_A\lvev q_Ae^{G'^A}|\Phi'^A|^2e^{2q_A\[U + \sum_Bb_A(L)\Sigma'^A\]}
\rvev e^{2q_Ah(L)} = {\dx\over2}L, \label{vevs}\eeq
giving an equation for the functional $h$:
\beq \sum_Aq_A\CA e^{2q_Ah(L)} = {\dx\over2}L.\label{funct}\eeq
Now 
\bea k(L) &\to& \tk(L) = k(L) + \delta k(L), \quad \delta k(L) = 
\lvev K(\Phi)\rvev = \sum_A\CA e^{2q_Ah(L)}, \nonumber \\
 2Ls(L) &\to& 2L\ts(L) = 2Ls(L) + 2L\delta s(L), \quad 2L\delta s(L) = 
-{\dx}Lh(L).\label{lts}\eea
We have using (\ref{funct})
\beq {\pp\over\pp L}\delta k = 2h'(L)\sum_Aq_A\CA e^{2q_Ah(L)} = L\dx
h'(L) = - 2L{\pp\over\pp L}\delta s \eeq
so from (\ref{canon}) the Einstein condition \myref{aab}
is satisfied for $U = \Sigma'^A=0.$

Now the full K\"ahler potential is
\bea K &=& k + G + K(\Phi), \nonumber \\
K(\Phi) &=& \sum_A e^{G'^A + 2q_A\[U + h(L) + \sum_Bb_B\Sigma'^B\]}
|\Phi'^A|^2 \nonumber \\ &=& \delta k(L) +
\sum_A e^{2q_Ah}\CA\[2q_AU + \(\Sigma'^A + 2q_A\sum_Bb_B\Sigma'^B\)
\(1 + 2q_AU\)\]
\nonumber \\ & & + O(U^2,\Sigma^2,|\Phi^{A>n}|^2) 
\nonumber \\ &=& \delta k(L) + L\dx U + 
\sum_A\Sigma'^A\[\CA e^{2q_Ah}\(1 + 2q_AU\) + b_A\dx\(L + U/h'\)\]
\nonumber \\ & & + O(U^2,\Sigma^2,|\Phi^{A>n}|^2),\label{almost}\eea
where we used the relation
\beq 4h'(L)\sum_Aq^2_A\CA e^{2q_Ah(L)} = \dx \eeq
that follows from $L$-differentiation of (\ref{funct}).
The physical, uneaten chiral supermultiplets $\Theta'^A$ do not
mix with the vector field $U$.   Thus we require
\beq 2q_A\CA e^{2q_Ah(L)} + b_A(L)\dx/h'(L) = f(L)q_AB_A, \eeq
which eliminates the $O(\Sigma U)$ terms in (\ref{almost}) by
virtue of the condition (\ref{gb}). We now have 
\bea K &=& \tk + G + L\dx U + \sum_A\Sigma'^A\CA e^{2q_Ah}\(1 - 2q_ALh'\)
 + O(U^2,\Sigma^2,|\Phi^{A>n}|^2),\nonumber \\ 
S &=& \ts + {1\over2}\tG - {1\over2}\dx\[U + h + \sum_Ab_A\Sigma'^A\] 
\nnn &=&
\ts + {1\over2}\tG - {1\over2}\dx\(U + h\) + h'\sum_Aq_A\CA e^{2q_Ah}\Sigma'^A,
\nonumber \\ \tG &=& bG + \dx\Gx.
\label{links}\eea
As before, we perform a Weyl transformation to put the Einstein
term in canonical form, replacing the functionals $K(L,M),S(L,M)$ by
$\hK(\hL,M),\hS(\hL,M)$. Identifying $\Delta_A = \Sigma'^A$ in the
notation of Appendix \ref{wtr}, we have
\bea k^A &=& \CA e^{2q_Ah}\(1 - 2q_ALh'\) = \CA e^{2q_Ah} - 2Ls^A,
\quad k'^A = - 2Ls'^A,  \nonumber \\ 
0 &=& {\alpha_0\over\dx}\(k'^A + 2Ls'^A\) = \alpha^A = k^{AU} 
= s^{AU}  \nonumber \\ &=& \hK^{AU} + 2\hL\hS^{AU} =  k^{AU} + 
2\hL s^{AU}
+ {\dx\hL\over3\alpha_0}\alpha^A\alpha^U. \eea
Therefore there is no $U,\Sigma'$ mixing in the effective K\"ahler
potential~\cite{BGG89,BGG01} $\tK = \hK + 2\hL\hS$ in the new Weyl basis.
The terms linear in $U$ can be eliminated as before, and further Weyl
transformations can be made involving $\Sigma^2,\Phi^2,U^2$ to get a
canonical Einstein term up to quadratic terms, as in the preceding
subsection.  

To see the relation of our quasi-unitary gauge to the actual unitary gauge that
can be determined only when the dilaton and moduli \vev's are fixed
by supersymmetry breaking, consider the relation between $U$ and the 
original vector field $V$ in terms of the original fields $\Sigma^A$.
\bea h + U &=& V + \hSig, \quad \hSig = \Sigma - \sum_Bb_B(L)\Sigma'^B 
= \sum_Bf_B(L)\Sigma^B, \nonumber \\
f_B(L) &=& {1\over2Q}q_BB_B\[1 + 2\sum_Aq_Ab_A(L)\] - b_B =
{q_Bx^B(L)\over2\sum_Cq^2_Cx^C(L)}, \nnn x^A(L) &=& \CA e^{2q_Ah(L)}
\label{uv}.\eea
If $L$ is replaced by a constant c-number $\ell_0$, this is just the
redefinition needed to go to unitary gauge starting with the K\"ahler
potential
\beq K(\Phi) = \sum_Ax^A(\ell_0)e^{\Sigma^A + 2q_A\tilde V}, \quad V =
\tilde V + h(\ell_0),\quad \lvev\tilde V\rvev = \lvev\Sigma^A\rvev = 0.
\eeq
Thus once the dilaton and moduli are stabilized and replaced by their
\vev's $\ell_0,t^I_0$ one is automatically in unitary gauge, independently of
the choice of the parameters $B_A$ and the functional $f(\ell)$. 
This freedom in parameter space is presumably
related to the large vacuum degeneracy in the absence of other couplings.
In particular one can impose $b_B(\ell_0) = 0,$ $\hSig(\ell_0) = \Sigma$,
by setting $B_A = 2x^A(\ell_0)/f(\ell_0)$. This still leaves the functional
$f(L)$ undetermined, but it fixes the effective modular weights 
(\ref{modwts}) and the the corresponding modification (\ref{uprime}),
(\ref{links}) to the GS term.  In the case that $\hSig(\ell_0)\ne
\Sigma$, a further transformation on the chiral multiplets
\beq \Phi''^A = e^{-2q_A\[\hat\Theta(\ell_0) - \Theta\]}\Phi' =  
e^{-2q_A\hat\Theta(\ell_0)}\Phi, \quad \hat\Theta(L) = 
\sum_Bf_B(L)\Theta^B\eeq
takes us into true unitary gauge, with the term $-\dx\sum_Ab_A(\ell_0)G'^A/2$
in $S$ in (\ref{links}) interpreted as an additional correction to the
GS term $\tG$.

Applying the results of Appendix \ref{wtr} for $\Delta^A = 
|\Phi^{A>n}|^2$,
the effective K\"ahler potential for matter is
\bea \tK(\Sigma',\Phi') &=& \hK(\Sigma',\Phi') + 2\hL\hS(\Sigma',\Phi')
= \sum_{A= 1}^n x^A\[\Sigma'^A + {1\over2}
\(\Sigma'^A + 2q_A\sum_Bb_B\Sigma'^B\)^2\]\ddd
+ \sum_{A>n}e^{2q_Ah+G'^A}|\Phi'^A|^2
 + O\(\Sigma^3,\Sigma|\Phi^{A>n}|^2\).\label{matter}\eea
Expanding around the dilaton vacuum in unitary gauge (with $U=0$) we have
\bea \left. \hL\right| &=& \ell_0 + \hel, \quad b_A = O(\hel), \nnn
x^A &=& x^A(\ell_0) + 2q_Ah'x^A(\ell_0)\hel + O(\hel^2), \quad
\sum_A q_Ax^A\Sigma'^A = O(\hel\Sigma'), \nnn
\tK(\Sigma',\Phi') &=& \sum_{A= 1}^n x^A(\ell_0)\[\Sigma'^A + {1\over2}
\(\Sigma'^A\)^2\]\ddd + \sum_{A>n}e^{2q_Ah+G'^A}|\Phi'^A|^2
 + O\(\Sigma^3,\Sigma^2\hel,\Sigma\hel^2,
\hel\Sigma|\Phi^{A>n}|^2\).\label{umatter}\eea
There is no kinetic mixing of the dilaton with the 
{\it D-moduli.}\footnote{The massless modes associated
with flat directions of the D-term part of the scalar
potential were referred to by this term in our previous
work \cite{gg1}.  Such moduli are a generic feature
of supersymmetric field theories \cite{BDFS}.}
The only
effect of the term linear in $\Sigma'$ is a slight modification of the
K\"ahler potential for the moduli:
\beq \del G_{I\bar J} = \sum_{A= 1}^n x^A(\ell_0)G'^A_{I\bar J}.\eeq
Once the T-moduli are fixed at their \vev's $t^I_0$, chiral D-moduli
can be defined as
\beq 
D^A = \widehat{\Theta}'^A + {1\over2}G'^A_I(t^I_0)\hat t^I, \quad
\Sigma'^A = D^A+ \bar D^A + O\([\hat t^I/t^I_0]^2\),
\eeq
where $t^I = t^I_0 + \hat t^I$.
When supersymmetry breaking is included~\cite{GG02c} via the
condensation
model of~\cite{bgw2}, $t_0^I = 1$ in reduced Planck units.  The
$D^A$ remain massless~\cite{BDFS,gg1}
in the absence of superpotential couplings.

The $U$ mass is given by
\beq m^2_U = {g^2_X\over2}\[\hK^{UU}(\ell_0) + 2\ell_0\hS^{UU}(\ell_0)\] =
{\dx g^2_X\over2}\lvev\({1\over h'} + {\ell_0\alpha_0\over3}\)\rvev
\label{vmass},\eeq
where $g_X$ is the effective \ux\, gauge coupling constant that will
be made explicit below, and $\alpha_0$ is defined in \myref{azd}.  This
reduces to the result found in~\cite{GG02a}
for the case of a single \vev\, with $h' = 1/2qL$.

To determine $g_X$ we start with the Yang Mills Lagrangian 
\beq
\L_{YM}= -{1\over8g^2}\superint {E\over R}(\W \W)_V, \quad \W_V = -
{1\over4}\chiproj\D V,\label{lymv}\eeq 
where $g$ is the renormalized
\ux\, coupling constant at the $U$ mass scale: $g^{-2} = s(\ell_0) +
\del_g$, with $\del_g$ containing corrections from $\L_Q + \superint
EL\tG$.  As noted in~\cite{GG02a}, the shift \myref{lts} in
$s(\ell_0)$ is canceled by the shift in $- \dx\superint ELV$ due to
the shift $h(\ell_0)$ in $V$.  However as discussed in~\cite{GG02a} we
have to include the various field redefinitions made in extracting the
Yang-Mills Lagrangian for $U$ 
\beq 
\L_{YM}= -{1\over8g_X^2}\superint
{E\over R}(\W \W)_U, \quad \W_U = - {1\over4}\chiproj\D
U.\label{lymu}
\eeq 
We have 
\beqa 
V &=& U + h(L) + \cdots = U + h(e^{\Del
K/3}\hL) + \cdots \nnn
&=& U\[1 + h'(\ell_0){\ell_0\alpha_0(\ell_0)\over3}\]
+ h(\hL) + \cdots, 
\eeqa
where the ellipses represent higher order terms that
are negligible for our purposes.  This gives, following \myref{kja},
\beq 
(\W \W)_V = \lvev\(1 +
h'{\hL\alpha_0\over3}\)^2\rvev(\W \W)_U + \cdots.
\eeq 
Thus \beq g_X = \lvev{3g\over3 +
h'\alpha_0\hL}\rvev , \quad m^2_U = \lvev{3g^2\dx\over h'(6 +
2h'\alpha_0\hL)}\rvev , \eeq which again reduces to the result
of~\cite{GG02a} in the case of a single \vev.  As discussed
in~\cite{GG02a} and the previous subsection, the operator \myref{lymv}
contains additional, higher dimensional operators, some of which are
linear in $U$. When $U$ is integrated out this leads to new operators
of very high dimension in the effective Lagrangian.  In the multi-vev
case there are additional operators due to the presence of the term
$\sum_Bb_B\Sigma^B$ in \myref{uv}.  These contributions will be of yet
higher dimension since $\lvev b_B\rvev = \lvev \Sigma^B\rvev = 0.$


\mysection{Generalized Gauge Group}
\label{ggg}
Next we consider the case of $m$ \uone's that are broken by $n$ 
scalar \vev's. In this case supersymmetry is unbroken provided
($\del_{aX}=0$ for $a \not= X$):
\beq 
\sum_{A=1}^n x^Aq_A^a = {\dx\ell\over2}\del_{aX}, \quad a = 1,
\ldots, m, \quad x^A(\ell) =\lvev K_{(A)}\rvev_{\ell},\label{da=0}
\eeq
which has a solution provided $n\ge m$ with $m$ of the
$n$ vectors $q^A_a$ linearly independent.  As usual, we will
promote (\ref{da=0}) to a superfield relation.    The two subsections
below closely parallel those in Section \ref{gfc} for
a single $U(1)$.

\subsection{Minimal Scalar Vevs}
We define a ``minimal'' set of scalar \vev's to mean a minimal set
of $n=m$ $x$'s that satisfy (\ref{da=0}). In this case
the physical spectrum is just $m$ massive vector superfields, and
it is straightforward to go to unitary gauge.
For $n=m$ (\ref{da=0}) has a unique inverse:
\beq
x^A(\ell) =  {\dx\ell\over2}Q^A_X, \qquad
x^A(L) =  {\dx L\over2}Q^A_X,
\label{invertx}
\eeq
where the matrix $Q^A_a$ is the inverse of $q^a_A$:
\beq \sum_aQ^A_aq^a_B = \del^A_B,\quad
\sum_Aq^a_AQ^A_b = \del^a_b.\label{qmatrix}\eeq
If $\lvev\Phi^A\rvev\ne0$ we can write
\beq \Phi^A = \xi_A^{1\over2}e^{\Theta^A},
\quad K_{(A)} = |\Phi^A|^2e^{G^A + 2\sum_aq^a_AV_a} = 
\xi_Ae^{\Theta^A + \bTh^A + G^A + 2\sum_aq^a_AV_a},\eeq
where $\xi_A$ is a positive real constant.  Then following the procedure
of section 2.1 we make a gauge transformation and a sequence of
field redefinitions:
\bea 
U'_a &=& V'_a + G_a = V_a + \Theta_a + \bTh_a + G_a =
U_a + h_a(L), \quad \lvev U_a\rvev \equiv 0,
\nonumber \\ 
Y_a &=& {1\over2}\sum_AQ^A_aY^A, \quad 
Y^A = 2\sum_aq^a_AY_a,\quad Y = \Theta,G,\label{uaredef}
\eea
giving
\beq 
K_{(A)} = \xi_Ae^{2\sum_aq^a_AU'_a} = 
\xi_Ae^{2\sum_aq^a_A[U_a + h_a(L)]} = x^A(L)e^{2\sum_aq^a_AU_a}. 
\eeq
We use (\ref{qmatrix}) to solve for the functionals $h_a(L)$:
\beq  
h_a(L) = {1\over2}\sum_AQ^A_a\ln(x^A/\xi_A) = 
{1\over2}\sum_AQ^A_a\ln(\dx Q^A_XL/2\xi_A).
\eeq
The constant parameters $\xi_A$ have no physical significance and just
reflect the fact that we can always make global $U(1)$ transformations
with constant parameters.  They drop out of 
\beq\del k(L) = \left.\sum_A K_{(A})\right|_{U_a=0} 
= {L\over2}\sum_A\dx Q^A_X, \eeq 
and we have
\bea
\del s(L) &=& {1\over2} \sum_A Q^A_a \ln(\dx Q^A_X L/ 2 \xi_A) =
\sum_AQ^A_a\[\ln L + \ln(\dx Q_X^A/2\xi_A)\], \nonumber \\
\superint\(L + \Omega\)\del s(L)
&=& {1\over2}\sum_AQ^A_a\superint\ln L\(L + \Omega\), \eea
by the linearity conditions.  So the $\xi_A$ drop out of
$\L_{GS} + \L_Q$; a natural choice is $\xi_A = \dx Q_X^A/2$.
As in the case of a single \ux, the modified functionals
\beq \tk(L) = k(L) + \del k(L), \quad \ts(L) = s(L) + \del s(L),\eeq
satisfy the Einstein condition (\ref{canon}):
\beq  \tk'(L) + 2L\ts'(L) = \del k'(L) + 2L\del s'(L) = 0,\eeq
and we have, instead of \myref{mqw},
\beq \tG = bG + {\dx\over2}\sum_AQ_X^AG^A = 
\sum_Ig^I\(b + {\dx\over2}\sum_AQ_X^Aq_I^A\).\eeq
Following section 2.1 we perform a Weyl transformation such that
in the transformed basis (\ref{ech}) of Appendix \ref{wtr} is 
satisfied.  As before, this eliminates the leading
order terms linear in $U$.
The quadratic term determines the vector boson mass matrix; as in 
Sections \ref{multvevs} we have to take into account the modification
of the effective coupling constant $g_U$, which is now a coupling
matrix, that is generated by the various field redefinitions.
We have
\bea
V_a &=& U_a + h_a(L) + \cdots 
= U_a + h_a(e^{\Del K/3}\hL) + \cdots \nnn
&=& U_a + h_a(\hL) + h'_a{\alpha_0\over3}\hL U_X + \cdots 
= U_a + h_a(\hL) + {\alpha_0\over6}\sum_BQ_a^BU_X + \cdots \nnn
&=& \mu_{ab}U^b + h_a(\hL) + \dots, \nnn
\sum_a(\W_V^a)^2 &=& \sum_{a,b,c}\mu_{ab}\mu_{ac}\W^b_U\W^c_U + \cdots
= \sum_d(\W^d)^2 + \cdots, 
\label{dv} 
\\
\W^b_U &=& \nu^{bd}\W_d, \quad
\sum_b\mu_{ab}\nu^{bd} = \del_a^c,\nonumber \\
\mu_{ab} &=& \del_{ab} + Q_a\del_{bX}, \quad \nu^{bd} = 
\del_{bd} - {Q_b\del_{dX}\over1 + Q_X}, 
\quad Q_a = {\alpha_0\over6}\sum_BQ^B_a.
\eea
The canonically normalized fields are $g_a^{-1}\W_a$,
with $g_a$ the renormalized \ua\, coupling constant
at the $U$ mass scale.  The squared mass
matrix $m^2_{ab}$ is given by
\bea (g_ag_b)^{-1}m^2_{ab}\W^a\W^b &=& {1\over2}\hK_{cd}\W^c_U\W^d_U 
= \nu^{ca}{1\over2}\hK_{cd}\nu^{db}\W_a\W_b, \quad 
\hK_{cd} = k^{cd} + {\alpha_0\over3}\dx\hL\del_{cX}\del_{dX}, \nonumber \\
k^{cd} &=& 4\sum_Aq^c_Aq^d_Ak^A = 2\sum_Aq^c_Aq^d_AQ^A_X\dx\hL.\eea
Putting everything together gives 
\beq m^2_{ab} = {g_ag_b\over2}\lvev k_{ab} 
- {\del_{aX}\del_{bX}\alpha_0\dx\hL
\over3(1 + Q_X)}\rvev,\label{massua}\eeq
which reduces to the result of~\cite{GG02a} 
in the case of only \ux\, with just one
scalar \vev: $k_{ab} = 2q\dx\hL,\;Q_X = \alpha_0/6q$. 
As before, the ellipses in (\ref{dv})
represent terms that generate higher dimension
operators in the effective theory below the $U$ mass scale.

\subsection{Nonminimal Case}
This is a straightforward generalization of the case discussed in 
Section \ref{multvevs}.
We first go to quasi-unitary gauge by setting
\bea U'_a &=& V_a + \Sigma_a, \quad \Phi'^A =
e^{-2\sum_aq^a_A\Theta_a}\Phi^A, \quad \Sigma_a = \Theta_a + \bTh_a +
G_a, \nonumber \\ \Theta_a &=&
{1\over2}\sum_{A,b}q^b_AM^{-1}_{ba}B_A\Theta^A, \quad G_a =
{1\over2}\sum_{A,b}q^b_AM^{-1}_{ba}B_A G^A, \nonumber \\ M_{ab} &=&
\sum_Aq^a_Aq^b_AB_A,\qquad a,b = 1\ldots m .\label{ugauge2}\eea
$\Phi'^A$ is chiral with modular weight 
\beq
q'^A_I = q^A_I -
2\sum_aq^a_Aq^a_I, \quad q^a_I =
{1\over2}\sum_{A,b}q^b_AM^{-1}_{ba}B_A q_I^A,\label{shifts}
\eeq 
and $U_a$ is modular invariant.  Without the $G_a$-terms, (\ref{ugauge2})
is a gauge transformation; $\Theta_a$ drops out of the Lagrangian.  If
$C_A\ne0$ for $A = 1,\ldots,n$, with $C_A$ defined as in \myref{defca},
\beq \Phi'^A =
C_Ae^{\Theta'^A}, \quad\Theta'^A = \Theta^A -
2\sum_aq^a_A\Theta_a.\eeq 
Only $n-m$ of the chiral fields
$\Theta'^A$ are linearly independent, where $m$ is the number of
linearly independent $\Theta_a$:
\beq \sum_Aq^a_AB_A\Theta'^A = 0\quad\forall a.\eeq 
The real fields $\Sigma'^A = \Theta'^A + \bTh'^A + G'^A$ have
vanishing $vev$'s and satisfy
\beq \sum_Aq^a_AB_A\Sigma'^A = 0\label{gb2}.\eeq
Now set 
\beq U'_a = U_a + h_a(L) + \sum_Bb_{aB}\Sigma'^B,
\quad \lvev U_a\rvev = 0.\eeq
Using the notation introduced above, we have
\bea \delta k(L) &=& \sum_A\CA \prod_ae^{2q_A^ah_a}, \nonumber \\
{\pp\over\pp L}\delta k(L) &=& 2\sum_A\CA
\sum_bq^A_bh'_b\prod_ae^{2q_A^ah_a}\nnn &=&  2\sum_bh'_b\sum_A\CA
q_A^b\prod_ae^{2q_A^ah_a} = h'_X\dx L, \label{delk}\eea
because the vanishing of D-terms requires
\beq \sum_A\CA q_A^b\prod_ae^{2q_A^ah_a} = \delta_{bX}{\dx\over2}
L\label{d=0}.\eeq
In addition
\beq 2L\delta s = - \dx L h_X, \quad
2L{\pp\over\pp L}\delta s = - \dx L h'_X = - {\pp\over\pp L}\delta
k,\label{dels}\eeq
and the Einstein condition \myref{aab} is again satisfied for
$U=\Sigma=0$. The K\"ahler potential for matter is
\bea
K(\Phi) &=& \sum_A e^{G'^A + 2\sum_aq^a_A
\[U_a + h(L)_a + \sum_Bb_{aB}\Sigma'^B\]}|\Phi'^A|^2
\nonumber \\ 
&=& \delta k(L) + L\dx U_X
+ \sum_A\Sigma'^A\(x^A + b_{XA}\dx L\) \nnn
&& + 2\sum_{A,a}\Sigma'^AU_a
\(q^a_Ax^A + 2\sum_{B,b}b_{bA}x^Bq^b_Bq^a_B\)
+ O(U^2,\Sigma^2,|\Phi^{A>n}|^2), \nnn
x^A &=& \CA e^{2\sum_bq^b_Ah_b}.
\eea
To eliminate $U,\Sigma'$ mixing we impose
\bea f_a(L)q^a_AB_A &=& q^a_Ax^A + 2\sum_{B,b}b_{bA}x^Bq^b_Bq^a_B,
\quad b_{aA} = {1\over2}\sum_bq^b_AN^{-1}_{ab}\(f_bB_A - x^A\),
\nonumber \\ N_{ab} &=& \sum_Bx^Bq^a_Bq^b_B, \quad \sum_Ab_{aA}\Sigma'^A
=  - {1\over2}\sum_bN^{-1}_{ab}x^Aq^b_A\Sigma'^A \equiv 
\sum_A\hb_{aA}\Sigma'^A,\eea
and we proceed as before with a Weyl transformation to make the
Einstein term canonical to quadratic order in $U,\Sigma,\Phi^{A>n}$.  
We now have
\bea 
K &=& \tk + G + L\dx U_X + \sum_Ak^A\Sigma'^A
+ O(U^2,\Sigma^2,|\Phi^{A>n}|^2), \nonumber \\
S &=& \ts + {1\over2}\tG 
+ {1\over2}\dx\(U_X + \sum_Ab_{XA}\Sigma'^A\), 
\quad \tG = bG + \dx\Gx, \nonumber \\
k^A &=& x^A + \hb_{XA}\dx L = x^A - 2Ls^A, \quad
k'^A = 2x^A\sum_aq^a_Ah'_a + \hb_{XA}\dx - 2Ls'^A.
\eea
Differentiation of (\ref{d=0}) with respect to $L$ gives
\beq 2\sum_ah'_aN_{ab} = \del_{bX}{\dx\over2}, \quad 2h'_a = N^{-1}_{aX}
{\dx\over2}, \quad \dx\hb_{XA} = - {\dx\over2}\sum_aN^{-1}_{aX} =
- 2x^A\sum_aq^a_Ah'_a. \eeq
So we have for $\DA = \Sigma'^A$
\beq 
k'^A + 2Ls'^A = 0 = \alpha^A, 
\eeq
and as in Section \ref{multvevs} the $U_a$ remain decoupled from the 
$\Sigma'^A$ in the Weyl transformed basis, and there are no terms
linear in $U$ in this basis. Following (\ref{uv}) we write
\bea h_a + U_a &=& V_a + \hSig_a, \quad \hSig_a = 
\Sigma_a - \sum_Bb_{aB}(L)\Sigma'^B 
= \sum_Bf_{aB}(L)\Sigma^B, \nonumber \\
f_{aB}(L) &=& {1\over2}\sum_bq^b_BB_B\(M^{-1}_{ab} + 2\sum_{Ac}q^c_A
M^{-1}_{cb}b_{aA}\) - b_{aB} =
{1\over2}\sum_bN^{-1}_{ab}q^b_Bx^B\label{uv2}.\eea
This again is the required redefinition of the vector field
at the dilaton vacuum: $L\to \ell_0$, and corresponds to the
true unitary gauge provided 
\beq \lvev B_{aA}\rvev = 0, \quad B_A = x^A(\ell_0)/f(\ell_0), \quad
f_a(\ell_0) = f(\ell_0),\eeq
which requires $f_a$ independent of $a$ up to terms that vanish
in the vacuum.

We set the massive $U_a$'s to zero 
to obtain the effective low energy theory.
As in Section \ref{multvevs} we can expand around 
the vacuum values of the dilaton and T-moduli to obtain 
the effective K\"ahler potential for matter: 
\bea \tK(\Sigma'^A,\Phi'^A) &=& \del G(T,\bar T) + \sum_{A= 1}^n x^A(\ell_0) 
\(D^A + \bar D^A\)^2 + \sum_{A>n}e^{2q_Ah+G'^A}|\Phi'^A|^2  
\ddd + O\(\hat\phi^3\),\label{umatter2}\eea 
where $\hat\phi$ is any field with vanishing \vev, and 
we used \beq \sum_A q^a_Ax^A\Sigma'^A = O(\hel\Sigma').\eeq 
The $n-m$ massless D-moduli $D^A$ and the shift 
$\del G_{I\bar J}$ in the T-moduli metric are defined as in 
subsection \ref{multvevs}.  

Finally, to determine the $U$ mass
matrix, we proceed as in the previous subsection:
\bea
V_a &=& U_a + h_a(L) + \cdots = 
U_a + \lvev h'_a{\alpha_0\over3}\hL\rvev U_X + 
h_a(\hL) + \cdots \nnn
&=& \mu_{ab}U^b + h_a(\hL) + \dots, \nonumber \\
\sum_a(\W_V^a)^2 &=& \sum_{a,b,c}\mu_{ab}\mu_{ac}\W^b_U\W^c_U + \cdots
= \sum_d(\W^d)^2 + \cdots, \nnn
\W^b_U &=& \nu^{bd}\W_d, \quad
\sum_b\mu_{ab}\nu^{bd} = \del_a^c,\nonumber \\
\mu_{ab} &=& \del_{ab} + Q_a\del_{bX}, \quad \nu^{bd} = 
\del_{bd} - {Q_b\del_{dX}\over1 + Q_X}, 
\quad Q_a = {\alpha_0\over6}h'_a, \label{dv2}\eea
giving 
\beq m^2_{ab} = {g_ag_b\over2}\nu^{ca}
\hK_{cd}\nu^{db} = {g_ag_b\over2}\lvev 
k_{ab} - {\del_{aX}\del_{bX}\alpha_0\dx\hL
\over3(1 + Q_X)}\), \eeq
which reduces to the result \myref{massua}
in the minimal case with $h'_a = \sum_AQ^A_a/\hL$.
As in Section \ref{multvevs}, the higher dimension operators implied
by the ellipses in (\ref{dv2}) include terms arising from 
$\sum_Ab_{aA}\Sigma^A$.

\mysection{Superpotential}
\label{spo}
Now we want to address linear couplings to heavy
fields that may appear in the superpotential, and
how these are eliminated by superfield redefinitions.
We will also have superpotential terms that
give masses to some more chiral multiplets.  
The $vev$'s have to be in F-flat directions,
so the superpotential has to be at 
most\footnote{In actuality, the superpotential can be more than
linear in $\Phi^A$ in terms that are of 
dimension greater than 3, i.e. in any term that
has at least 2 fields with vanishing $vev$'s.} 
linear in $\Phi^A$ if $C_A\ne0$.  
The superpotential terms are of the form
\beq \prod_A\Phi^Af(T) \eeq
where $f(T)$ makes the expression modular covariant.  
Now $\prod_A\Phi^A$ is invariant
under all the $U(1)$'s: $\sum_Aq^a_A = 0$, 
so when we make the gauge 
transformation (\ref{ugauge2})
the total modular weight doesn't change and the 
superpotential remains modular covariant, as 
can be checked. Now suppose we have a term 
\beq \Phi_1\Phi_2\Phi_3f(T), \quad C_1\ne 0.\eeq
$\Phi_2,\Phi_3$ combine to form a massive supermultiplet.  
There might also be terms linear in these:
\beqa 
W &\ni& \Phi_1\Phi_2\Phi_3f(T) + \Phi_2w_3(\Phi) + \Phi_3w_2(\Phi)
= \Phi_1\Phi'_2\Phi'_3f(T)
- (\Phi_1f)^{-1}w_2w_3, \nnn
\Phi'_i &=& \Phi_i + (\Phi_1f)^{-1}w_i(\Phi).
\label{iwy}
\eeqa
The last term is dimension 4 since $w^i \sim (\Phi)^2$ 
and by assumption $\lvev w_i\rvev\ne0$,
so for most purposes we can drop it and set $\Phi'_{2,3} = 0$ 
in the superpotential.  However, the last term
may be of interest if it generates highly suppressed
operators that are otherwise forbidden.

For example suppose
\beq
w_2 = \lambda^2_{ij} Q^i Q^j, \qquad w_3 = \lambda^3_{ij} Q^i L^j,
\eeq
where $Q^i \; (i=1,2,3)$ are the three generations
of quark doublet superfields and $L^i \; (i=1,2,3)$
are the lepton doublet superfields.
We identify $\Phi_2 \equiv D^c$, $\Phi_3 \equiv D$,
a gauge-vector pair of color-triplet, $SU(2)_L$
singlet, chiral superfields.  Thus we
obtain from \myref{iwy} the effective superpotential
operator
\beq
- (\Phi_1f)^{-1}w_2w_3 = - (\Phi_1f)^{-1} \lambda^2_{ij} 
\lambda^3_{k \ell} Q^i Q^j Q^k L^\ell
\eeq
that mediates nucleon decay.  In contrast to the minimal
GUT case, the couplings $\lambda^2_{ij}$ and $\lambda^3_{ij}$
have no reason to be hierarchically small for the light
quark generations, since these are not the Yukawa
couplings that give masses to these quarks.  Thus,
even if the effective vector mass for the $D,D^c$
pair is of the order of the usual colored Higgs scale,
we can exceed proton decay limits by several orders
of magnitude.  To further address such issues requires a model
dependent analysis of the string scale couplings 
of MSSM supermultiplets to exotic supermultiplets
and the flat directions that yield the effective
vector mass terms for exotic quarks.

\mysection{Conclusions}
\label{con}
We have considered the case of several scalar fields, charged under a
number of $U(1)$ factors, acquiring vacuum expectation values due to
an anomalous $U(1)$.  We have demonstrated how to make redefinitions
at the superfield level in order to account for tree-level exchange of
massive vector superfields in the effective supergravity theory of the light
fields in the supersymmetric vacuum phase.  Our approach has built
upon previous results that we obtained in a more elementary case.  We
found that the modular weights of light fields are typically shifted
from their original values, allowing an interpretation in terms of the
preservation of modular invariance in the effective theory.  We have
addressed the subtleties in defining unitary gauge, associated in part
with the noncanonical K\"ahler potential that occurs in modular
invariant supergravity.  Further complications arise from the role of
the dilaton as the order parameter in the (most realistic) case where
the vacuum is degenerate in \uone-charged scalar space (D-moduli
space).  We have discussed the effective superpotential for the light
fields and have noted how proton decay operators may be obtained when
the heavy fields are integrated out of the theory at the tree-level.

We still need to include spontaneous symmetry breaking
by gaugino condensation in a hidden sector and the
related soft supersymmetry breaking phenomenology.
Work in this direction will be presented in
a future publication \cite{GG02c}.  A related issue,
that will also be taken up in \cite{GG02c},
is the stabilization of the dilaton $\ell$
in the presence of matter fields with large
\vev's.  
In \cite{Gie02a} it was noted that all models in
the class studied there suffer from a T-moduli
mass problem.  Stabilization of the T-moduli
through an effective theory of gaugino
condensation will allow us to address how the
moduli masses may change in the presence of a
\ux\ factor; indeed, we will show that the mass
{\it is} modified and that this may ameliorate the
moduli problem discussed in \cite{Gie02a}.
It also remains to be studied how the required
couplings to hidden matter condensates will
stabilize the D-moduli.  This has been touched
on in a previous letter \cite{Gie02b}, but a full-fledged
analysis where tree-level exchange has been taken
into account needs to be performed.  This too will
appear in future work \cite{GG02c}.  Finally,
it is important to understand how the presence of
large \vev's could effect other phenomenological
aspects of semi-realistic models.

\vspace{0.20in}

\noindent {\bf \Large Acknowledgments}

\vspace{5pt}

\noindent 
 This work was supported in part by the Director, Office of Science,
Office of High Energy and Nuclear Physics, Division of High Energy
Physics of the U.S. Department of Energy under Contract
DE-AC03-76SF00098 and in part by the National Science Foundation under
grant PHY-0098840.

\myappendix

\mysection{Canonical Einstein Normalization}
\label{ecd}
Of chief concern in our considerations is the
maintenance of the canonical normalization for the Einstein
term---concurrent to field redefinitions.  
Therefore we lay out a general prescription
for determining the necessary {\it Einstein condition}
from $\L$ rewritten in a new field basis.

The relevant part of the Lagrangian is \myref{oit}.  We define $M$ to
stand collectively for the fields that are to be
regarded as independent of $L$ in a given basis.  We
then define the functional $S$ by the identification
\beq
\tL \equiv E[-3 + 2L S(L,M)].
\label{jut}
\eeq
The Einstein condition holds provided
\beq
\( {\p K \over \p L}\)_M + 2 L \( {\p S \over \p L}\)_M = 0.
\label{aab}
\eeq
Here the subscripts on parentheses
instruct us to hold constant
under differentiation the fields denoted
collectively by $M$.

\mysection{Weyl Transformation}
\label{wtr}
First we fix to unitary gauge such that we
have a basis of fields written in terms of $L$
and modular invariant real superfields 
$\DA$ that satisfy $\myvev{\DA}=0$.  For instance
in the text the $\DA$ stand collectively for unitary
gauge vector multiplets $U_a$ with vanishing \vev's, 
the modular invariant composite superfields $X_A$, 
and, for $\myvev{X_A}\ne 0$, the superfields
$\Sigma_A = \ln(X_A/\myvev{X_A})$.
We can quite generally write the K\"ahler
potential as a power series in the superfields
with vanishing \vev's:
\beq
K(L,g^I,\Del) = \tk(L) + G + \sum_A k^A(L) \DA 
+ \half \sum_{AB}k^{AB}(L) \DA \DB + \ord{\Del^3}.
\label{wke}
\eeq
Similarly we can write the functional $S$ defined
in Appendix \ref{ecd} as
\beq
S(L,g^I,\Del) = \ts(L) + \half \tG + \sum_A s^A(L) \DA 
+ \half \sum_{AB} s^{AB}(L) \DA \DB + \ord{\Del^3}.
\label{wse}
\eeq
The functionals $G$ and $\tG$ are defined in
the main text and are independent of $L$ and the
fields $\DA$.  We also assume the identity, shown
in the text to hold,
\beq
\tk'(L) + 2 L \ts'(L)=0.
\label{tci}
\eeq

To put the Einstein term in canonical 
form for the terms involving the $\DA$ fields,
we make a Weyl transformation 
\beq
K \equiv \hK + \Dk
\label{ksh}
\eeq
such that
\beq 
E = e^{-\Dk/3} \hE, \qquad L = e^{\Dk/3}\hL,
\label{wyl}
\eeq
with $\Dk$ an $\ord{\Del}$ functional with assumed form
\beq
\Dk(\hL,\Del) = \sum_A \alpha^A(\hL) \DA 
+ \sum_{AB} \beta^{AB}(\hL) \Del_A \Del_B + O(\Del^3).
\label{kex}
\eeq
Thus we pass to a new superfield basis $(\hL,g^I,\Del)$.
From \myref{jut} the Einstein condition is restated
in terms of a new functional $\hS(\hL,g^I,\Del)$ given by
the identification
\beq
2 \hL \hS(\hL,g^I,\Del) \equiv 
2 \hL \left. S(L,g^I,\Del) \right|_{L = e^{\Dk/3}\hL} 
+ 3\(1 - e^{-\Dk/3}\).
\label{shd}
\eeq

The field redefinition (\ref{wyl}) assures that the linearity condition for $L\to\hL$
is not modified.\footnote{As discussed in the main text,
however, the redefinitions of vector superfields lead
to a reinterpretation of the chiral field strengths
in the new superfield basis.  Similarly, the Chern-Simons
superfield $\Omega$ used here must also be reinterpreted
in the new superfield basis.}
This can be seen by first leaving $L$ unconstrained and writing
\bea \tL &\to& \tL' = \tL - E(S + \S)\(L + \Omega\), \label{tlp}\eea
where $\Omega$ is the Chern-Simons superfield: 
$\(\DbDb - 8R\)\Omega = \W^\alpha \W_\alpha$.
Under a Weyl transformation: $E\to X E = \hE$, $\Omega\to X^{-1}\Omega = \hO$, 
$L\to X^{-1}L = \hL$, $E\Omega = \hE\hO$ and $EL = \hE\hL$ are Weyl invariant~\cite{adam}.
Then ({\ref{tlp}) takes the same form in 
terms of the hatted fields, and the equations
of motion for the chiral and anti-chiral
superfields $S,\S$ give the linearity conditions for $\hL$.

In the new basis and in terms of the functionals
$\hK$ and $\hS$ we require \myref{aab} to have
the canonical Einstein term:
\beq
\( {\p \hK \over \p \hL} \)_{g^I,\Del} 
+ 2 \hL \( {\p \hS \over \p \hL}\)_{g^I,\Del} = 0.
\label{ech}
\eeq
Using the definitions \myref{ksh} and \myref{shd} it
is straightforward to express \myref{ech} in terms
of the original functionals $K$ and $S$, as well
as the Weyl transformation functional $\Dk$:
\beq
{\p \hK \over \p \hL} + 2 \hL {\p \hS \over \p \hL}
= {\p L \over \p \hL} \[ K'(L) + 2 \hL S'(L) \]
- \({3 \over \hL} - {\p \Dk \over \p \hL} \)
\( 1 - e^{-\Dk/3} \).
\label{hty}
\eeq
Here, we use the shorthand notation
\beq
K'(L) \equiv \left. {\p K(L,g^I,\Del) \over \p L}
\right|_{L = e^{\Dk/3}\hL}
\eeq
and similarly for $S'(L)$.  We also note that 
\beq
{\p L \over \p \hL} = {\p L(\hL,\Del) \over \p \hL}
= e^{\Dk/3} \(1 + {\hL \over 3} {\p \Dk \over \p \hL} \)
\eeq
is easily computed using \myref{wyl}.

We next expand \myref{hty} in powers of $\Del$
and demand that it vanish at each order so that
\myref{ech} will hold in the new basis.  This
involves power series expansion in $\Delta$ of the $L$-dependent
coefficients appearing in \myref{wke} and \myref{wse}
corresponding to the expansion of the quantity
\beq
K'(L) + 2 \hL S'(L) = K'(\hL) + 2 \hL S'(\hL) 
+\hL \( e^{\Dk/3}-1 \) \[ K''(\hL) + 2 \hL S''(\hL) \] 
+ \ord{\Del^2},
\eeq
where we denote here and elsewhere below
\beq
K'(\hL) \equiv \left. K'(L) \right|_{L=\hL}, \qquad
K''(\hL) \equiv \left. K''(L) \right|_{L=\hL}, 
\qquad {\rm etc.}
\eeq

Since $K'(\hL) + 2 \hL S'(\hL)=\tk'(\hL) + 2\hL\ts'(\hL) = 0$ holds
identically with Eq.~\myref{tci}, 
we have that $K'(L) + 2\hL S'(L)$ is $\ord{\Del}$.
Thus the first nontrivial conditions that arise are
at the order of terms linear in $\Del$.  We obtain
\beq
{\p \hK \over \p \hL} + 2 \hL {\p \hS \over \p \hL} 
= \sum_A \left\{ k'^A(\hL) + 2 \hL s'^A(\hL) 
+ {\hL \over 3} \[\tk''(\hL) + 2 \hL \ts''(\hL) 
- {3 \over \hL^2} \] \alpha^A(\hL) \right\} \DA + 
\ord{\Del^2},
\label{wc2}
\eeq
where
\beq
k'^A(\hL) = \left. {dk^A(L) \over dL} \right|_{L=\hL}, \qquad
s'^A(\hL) = \left. {ds^A(L) \over dL} \right|_{L=\hL}, \qquad
{\rm etc.}
\eeq
We must choose each term in the sum of Eq.~\myref{wc2} to
vanish.  It is convenient to rewrite these constraints
in the following manner.  First note that 
differentiating Eq.~\myref{tci}
and then sending $L \to \hL$ yields
\beq
\hL \[ \tk''(\hL) + 2 \hL \ts''(\hL) \]
= - 2 \hL \ts'(\hL) = \tk'(\hL).
\label{kpq}
\eeq
Furthermore we introduce the functional $\alpha_0(\hL)$
defined in our previous article:
\beq
\alpha_0(\hL) \equiv {3 \dx \hL \over 3 - \hL \tk'(\hL) }.
\label{azd}
\eeq
Then the right-hand side of \myref{wc2} may be written
\beq
{\p \hK \over \p \hL} + 2 \hL {\p \hS \over \p \hL} 
= \sum_A \[ k'^A(\hL) + 2 \hL s'^A(\hL) 
-\dx {\alpha^A(\hL) \over \alpha_0(\hL)} \] \DA 
+ \ord{\Del^2}.
\eeq
To have the Einstein condition satisfied to $\ord{\Del}$
we require that the linear coefficients in \myref{kex}
be taken as
\beq
\alpha^A(\hL) = {1 \over \dx} \alpha_0(\hL)
\[ k'^A(\hL) + 2 \hL s'^A(\hL) \].
\label{adf}
\eeq

Once the $\hL$-dependent coefficients in \myref{kex}
have been determined from the requirement \myref{ech},
as has just been done to $\ord{\Del}$, then the functionals
$\hK(\hL,g^I,\Del)$ and $\hS(\hL,g^I,\Del)$ are completely
determined.  We define corresponding expansions in the
``small'' superfields $\DA$:
\beqa 
\hK(\hL,g^I,\Del) & \equiv & \tk(\hL) + G 
+ \sum_A \hK^A(\hL) \Del_A 
+ {1\over2} \sum_{AB} \hK^{AB}(\hL) \Del_A\Del_B 
+ O(\Del^3),
\label{que} \\
\hS(\hL,g^I,\Del) &=& \ts(\hL) + \half \tG
+ \sum_A \hS^A(\hL) \Del_A 
+ {1\over2} \sum_{AB} \hS^{AB}(\hL) \Del_A \Del_B 
+ O(\Del^3),
\label{quf}
\eeqa
It is useful to determine the $\hL$-dependent
coefficients to leading orders using the $\ord{\Del}$
results given above, e.g., Eq.~\myref{adf}.

Making the necessary expansions of the $L$-dependent
quantities that appear in \myref{wke} and \myref{wse}
to get to the basis $(\hL,g^I,\Del)$, we obtain
\beqa
\hK^A(\hL) &=& \( \third \hL \tk'(\hL)-1 \) \alpha^A(\hL)
+ k^A(\hL),
\\
2 \hL \hS^A(\hL) &=& 
\( 1 + {2 \over 3} \hL^2 \ts'(\hL) \) \alpha^A(\hL)
+ 2 \hL s^A(\hL),
\nnn
&=& \( 1 - \third \hL \tk'(\hL) \) \alpha^A(\hL)
+ 2 \hL s^A(\hL),
\eeqa
where we have used \myref{tci} in the last step.  Taking
\myref{adf} and \myref{azd} into account we rewrite
these as
\beqa
\hK^A(\hL) &=& k^A(\hL) - \hL \( k'^A(\hL) + 2 \hL s'^A(\hL) \),
\\
2 \hL \hS^A(\hL) &=& 2\hL s^A(\hL) + \hL\(k'^A(\hL) 
+ 2\hL s'^A(\hL) \).
\eeqa
We note the simplicity of the $\ord{\Del}$ contribution to
$\hK + 2 \hL \hS$:
\beq
\hK^A(\hL) + 2 \hL \hS^A(\hL) = k^A(\hL) + 2 \hL s^A(\hL).
\label{liw}
\eeq
Since $U_a$ is linear in $V_a$ we have
\beq k^{U_a} = k^{V_a} \equiv k^a, \quad s^{U_a} = s^{V_a} 
\equiv s^a. \eeq
Then it follows from \myref{sfid} that
\beqa
k^a(\hL) + 2 \hL s^a(\hL) &=& 0, 
\qquad \hL \( k'^a(\hL) + 2 \hL s'^a(\hL) \) 
= - 2 \hL^2 s^a(\hL) = k^a(\hL), \nnn
\hK^a(\hL) &=& 2 \hL \hS^a(\hL) = 0.
\eeqa
Thus the unwanted linear couplings in $U_a$ are automatically removed
by the Weyl transformation.

The $\ord{\Del^2}$ coefficient functionals are more
complicated, and involve the functional $\beta^{AB}(\hL)$
that should follow from \myref{ech} and 
\myref{hty} at $\ord{\Del^2}$
but was not explicitly determined above.  However,
for our purposes we need only the $\ord{\Del^2}$ contribution to
$\hK + 2 \hL \hS$, that as it turns out does not
depend on $\beta^{AB}(\hL)$.  We find:
\bea 
\hK^{AB} + 2\hL \hS^{AB} &=&
k^{AB}(\hL) + 2\hL s^{AB}(\hL)
+ \alpha^A \alpha^B
\[ {\hL^2 \over 9} \( \tk''(\hL) + 2 \hL \ts''(\hL) \) - \third \]  
\nnn & & + {\hL \over 3} \[ 
\( k'^A(\hL) + 2\hL s'^A(\hL) \) \alpha^B
+ \( k'^B(\hL) + 2\hL s'^B(\hL) \) \alpha^A \]
\nnn 
&=& k^{AB}(\hL) + 2\hL s^{AB}(\hL) 
+ {\dx \hL \over 3 \alpha_0} \alpha^A \alpha^B .
\label{qdw}
\eea 
In the second step we have exploited \myref{tci}, \myref{kpq}, 
\myref{azd} and \myref{adf} to simplify considerably.

In the theory defined by \myref{tlg} 
quadratic terms in the $\DA$ appear only through the combination
of functionals $\hK + 2\hL\hS$.  The quantity
given in \myref{qdw} is therefore the effective
metric for these supermultiplets.
Research in progress \cite{GG02c} shows
that when a gaugino condensate potential is added, 
quadratic terms in $\DA$ appear through
$\hK$ by itself, rather than in the 
combination $\hK + 2\hL\hS$.  However, this lone $\hK$
appears only in terms proportional to the squared 
condensate $|\myvev{\lambda \lambda}|^2$. 
We find that any nonvanishing \vev's $\myvev{\DA}$ 
are naturally of the same order (since they represent
a shift away from the supersymmetric vacuum
that was stable in the absence of gaugino condensation),
so we don't have to evaluate $\beta^{AB}$
because it only appears in these negligible terms.

\mysection{Generalized GS Mechanism}
\label{usi}
In this appendix we address the situation that
arises in Type I and Type IIB four-dimensional
$N=1$ superstring models; for example in \cite{IQ99}.
Here the anomaly matching that occurs in
the weakly-coupled heterotic models considered
above no longer holds and a more general
GS term is required.  The cancellation of \uone\ anomalies
results from couplings to two-forms from the {\it twisted} closed
string sectors, which we will denote $b_{(A)mn}$.
Since these two-forms are contained in linear multiplets
of the underlying theory \cite{ABD99},
the generalized GS mechanism can easily be incorporated into
the present formalism, as we now describe.

Suppose a modification of the GS counterterm Lagrangian
such that
\beq
\L_{GS} \ni -\half \sum_{a \in \G_X,A} c_{Aa} B_{(A)}^m v_{(a)m}.
\label{ggs}
\eeq
Here, $\G_X$ is a product of anomalous \uones\ and
$v_{(a)m}$ are the corresponding vector bosons.  The
one-form $B_{(A)}^m$ is a gauge-invariant (dual) field strength
obtained from coupling the two-form to a combination of
Chern-Simons three-forms for each of the simple factors
$\G_a$ in the full gauge group $\G$:
\beq
B_{(A)}^m = \half \e^{mnpq} \( \p_n b_{(A)pq} +
{2 \over 3} \sum_{a \in \G} \tc_{Aa}\Omega_{(a)npq} \).
\label{hyq}
\eeq
This is a straightforward generalization of the
coupling of a single two-form field strength to
Chern-Simons three-forms, as has been described
for instance in \cite{adam,GG99,BGG01}.  It then
follows from this definition that
\beq
\p_m B_{(A)}^m = \sum_{a \in \G} \tc_{Aa} (F \cdot \tilde F)_a .
\eeq
With this in mind it is easy to see that under
a gauge transformation acting on the anomalous
\uone\ vector bosons according to
$v_{(a)m} \to v_{(a)m} + \p_m \lambda_{(a)}$,
Eq.~\myref{ggs} shifts as (up to a total derivative)
\beqa
\delta \L_{GS} &=& -\half \sum_{a \in \G_X,A} c_{Aa} B_{(A)}^m 
\p_m \lambda_{(a)} 
= \half \sum_{a \in \G_X,A} c_{Aa} \lambda_{(a)} \p_m B_{(A)}^m \nnn
& = & \half \sum_{a \in \G_X,b \in \G} \hat c_{ab} \lambda_{(a)}
(F \cdot \tilde F)_b, \qquad \hat c_{ab} \equiv 
\sum_A c_{Aa} \tc_{Ab} .
\eeqa
The anomaly cancellation coefficients $\hat c_{ab}$
can then be matched to those obtained from the underlying
theory; e.g., any of the matrices enumerated in \cite{IRU99}.

The terms \myref{ggs} are obtained in the
case where the GS counterterm Lagrangian for the
anomalous \uones\ is given by
\beq
\L_{GS}^X = - \int E \sum_{a\in \G_X,A} c_{Aa} L_A V_a.
\label{las}
\eeq
Here $V_a$ are the vector superfields corresponding to the
anomalous \uones\ and $L_A$ are linear superfields
that arise in the twisted closed string sector.  These
linear multiplets are coupled to Chern-Simons superfields
in a manner implied by \myref{hyq}, which leads to modified
linearity conditions:
\beqa
\chiproj L_A &=& -(\W \W)_A, \qquad \chiprojb L_A = - (\Wb \Wb)_A,
\label{cic} \\
{[} \D_\alpha, \Dbadd ] L_A &=& 4 L_A G_{\alpha \adot}
+ 2 B_{A \alpha \adot} + 2 (\W_\alpha \Wb_\adot)_A, \\
(\W \W)_A & = & \sum_a \tilde c_{Aa} (\W \W)_a, \qquad
{\rm etc.}
\eeqa

Corresponding to \myref{las} it is necessary to generalize
\myref{lqb}:
\beq
B_b = \sum_I (b-b^I_b) g^I -  
\sum_{a\in \G_X} \hat c_{ab} V_a + f_b(L) .
\label{ona}
\eeq
It is easy to see that the anomalous shift generated by
\myref{ona} is canceled by \myref{las}.
For suppose $V_a \to V_a' + \half (\Theta_a + \bar \Theta_a)$
with $U(1)_a \in \G_X$.  We then have
\beq
\delta \L_Q = \int {E \over 16 R} 
\sum_{a \in \G_X,b \in \G} \hat c_{ab} \Theta_a (\W \W)_b + \hc
\eeq
On the other hand \myref{las} shifts according to
\beqa
\delta \L_{GS}^X &=& - \half \int E \sum_{a\in \G_X,A} 
c_{Aa} L_A \Theta_a + \hc \nnn
&=& \int {E \over 16 R} \sum_{a\in \G_X,A} 
c_{Aa} \Theta_a \chiproj L_A + \hc \nnn
&=& - \int {E \over 16 R} \sum_{a \in \G_X, b \in \G, A} 
c_{Aa} \tilde c_{Ab} \Theta_a (\W \W)_b + \hc \nnn
&=& - \int {E \over 16 R} \sum_{a \in \G_X,b \in \G} 
\hat c_{ab} \Theta_a (\W \W)_b  + \hc
\eeqa
so that $\delta \L_Q + \delta \L_{GS}^X=0$.

In other respects the effective Lagrangian can be 
formulated according to the approach described in \cite{Wu99}.
From the universal dilaton linear multiplet $L$,
other untwisted closed string linear multiplets $L_i$
and the twisted closed string linear multiplets $L_A$
we can form a linear combination corresponding to
each factor $\G_a$ of the gauge group:
\beq
L_a = \zeta_a^L  L + \sum_i \zeta_a^i  L_i 
+ \sum_A \zeta_a^A L_A .
\eeq
The coefficients are chosen such that the $L_a$
satisfy modified linearity conditions corresponding
to a coupling to the Chern-Simons superfields
of the factor $\G_a$:
\beqa
\chiproj L_a &=& -(\W \W)_a, \qquad \chiprojb L_a = - (\Wb \Wb)_a, \\
{[} \D_\alpha, \Dbadd ] L_a &=& 4 L_a G_{\alpha \adot}
+ 2 B_{(a)\alpha \adot}+ 2 (\W_\alpha \Wb_\adot)_a.
\eeqa
Following \cite{Wu99}, the $L$ dependent functionals
appearing in the main text can then be generalized:
\beq
k(L) \to \sum_a k_a(L_a), \qquad
2L s(L) \to \sum_a 2 L_a s_a(L_a).
\eeq
Further details may be found in \cite{Wu99}.

\end{document}